# Activité motrice des truies en groupes dans les différents systèmes de logement

Yannick Ramonet, Anaïs Tertre
Chambre d'agriculture de Bretagne. Février 2014

*Les modes de logement des truies en groupes se distinguent par les dimensions des cases et le système d'alimentation. Des observations ont été réalisées en élevages de production pour évaluer le niveau d'activité physique et les distances parcourues par les truies. De 50 à 500 m parcourus en 6 heures d'observation, le mode de logement impacte l'activité motrice des animaux.*

## Introduction

Les modes de logement pour les truies en groupes se différencient notamment par les dimensions des cases et la taille de groupes d'animaux (Ramonet et al., 2011). Les critères de réussite pour agencer les salles ou conduire les animaux en groupes ont fait l'objet de nombreux travaux et enquêtes.

La manière dont les truies en groupes se comportent selon le mode de logement est peu connue. Sont-elles davantage actives dans un système plutôt qu'un autre ? Comment utilisent-elles l'espace disponible ? Le type de sol, et la présence de litière notamment, engendre-t-il des comportements particuliers ?

Préciser le niveau d'activité des truies en groupes selon le mode de logement est important pour au moins deux raisons :

1) L'activité motrice des truies peut être impliquée dans les troubles locomoteurs observés chez des truies en groupes. On peut penser que plus une truie se déplacera, plus elle sera susceptible de développer une boiterie ou des lésions sur les pattes, et moins elle pourra bénéficier d'un temps de repos qui lui permettra de se remettre. Les problèmes locomoteurs des truies en groupes apparaissent très liés au mode de logement (Anil et al., 2007 ; Caille, 2012).

Dans une enquête récente en élevages, Cador (2013) montre ainsi que le caillebotis présente un facteur de risque supplémentaire par rapport à la litière. Le DAC conduit de manière dynamique ou statique est plus à risque que les systèmes bat-flanc et réfectoire-courette.

2) Par ailleurs, l'activité physique des truies induit un besoin énergétique supplémentaire. Le besoin énergétique d'entretien est doublé lorsque les truies sont debout par rapport à une position couchée (Noblet et al., 1994). Ce paramètre est intégré dans le modèle de simulation des besoins alimentaires InraPorc (Dourmad et al., 2008). La donnée de référence est que la truie est en position debout 240 minutes par jour. InraPorc permet de moduler le temps passé debout, et donc de calculer le besoin alimentaire de la truie. Le modèle est cependant difficile à ajuster compte tenu du peu de références disponibles. Les différences de comportement entre modes de logement sont documentés (Paboeuf et al., 2010), mais la comparaison entre études reste difficile compte tenu de la diversité des protocoles utilisés pour réaliser les observations.

L'objectif de la présente étude est de préciser, en élevages de production, le niveau d'activité motrice de truies logées dans des systèmes distincts

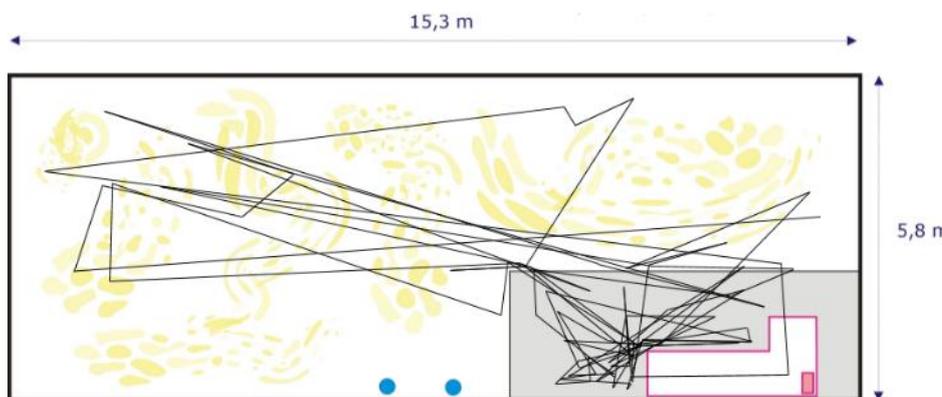

Parcours P1 :

Elevage 8 (crécom), DAC stable et litière

Tracé du parcours d'une truie lors de 6 h d'observations. La truie explore l'ensemble de la surface de la case.

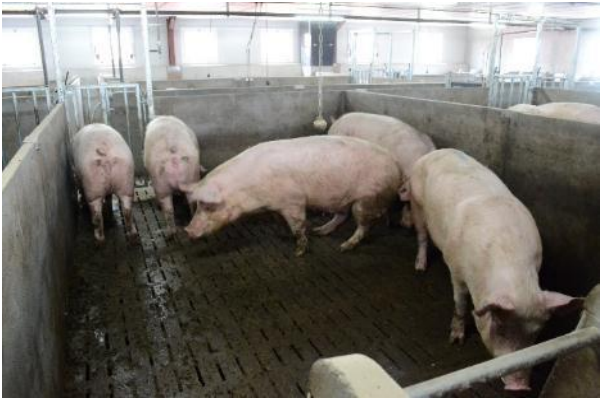 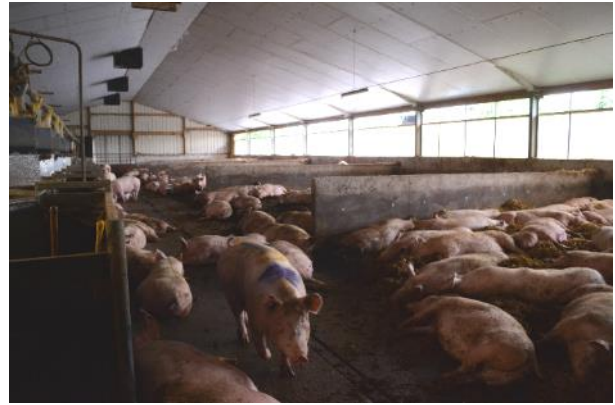

*Photo 1 : Contraste entre 2 modes de logements pour les truies en groupes. En groupe de 6 truies dans une case de 15 m² sur caillebotis (à gauche) ou en groupe dynamique de 280 individus dans une salle de 490 m² sur litière (à droite). Le mode de logement amène les truies à différencier leur activité motrice.*

# 1. Matériel et méthodes

## 1.1. Elevages

Neuf élevages ont été retenus, dont celui des Chambres d'agriculture de Bretagne à Crécom dans lequel 3 équipements différents sont observés et que l'on considèrera comme 3 élevages indépendants. Les élevages se distinguent par le mode de logement et le type de sol pour les truies en groupes (Tableau 1): bat-flanc/caillebotis (2 élevages) ; réfectoire-courette/caillebotis (1 élevage) réfectoire-courette/litière (3 élevages) ; DAC, groupe stable/caillebotis (2 élevages) ; DAC, groupe stable/litière (1 élevage) ; DAC, groupe dynamique /litière (1 élevage) ; DAC, groupe dynamique /caillebotis (1 élevage).

Dans chacun des élevages, 10 truies multipares d'une même bande, en milieu de gestation, sont observées. Elles sont sélectionnées au hasard dans la bande. Dans le cas du DAC, les truies observées sont toutes dans le même groupe. En revanche dans les systèmes de type réfectoire-courette et bat-flanc, les truies observées peuvent appartenir à plusieurs groupes différents.

## 1.2. Mesures et analyses statistiques

Les observations du comportement sont réalisées pendant 6 heures au cours de 2 journées consécutives selon la méthode du scan sampling.

Toutes les 4 minutes, l'emplacement de la truie, sa posture (debout-assise-couchée) ainsi que son comportement (alimentation-activité orale non alimentaire-manipulation de la paille-incativité-contact avec des congénères) sont reportés sur un plan du bâtiment. Les observations débutent au moment du repas ou de la séquence alimentaire du matin (bat-flanc ; réfectoire-courette ; DAC groupe stable) ou deux heures avant le début de la séquence alimentaire dans les 2 élevages avec DAC groupe dynamique pour lesquels les séquences alimentaires débutaient à 17h et 18h30.

Des macros développées sur un tableur Excel permettent de tracer le parcours de chaque truie et de calculer la distance parcourue.

Sur chaque période de 6 heures, le budget temps de la truie est calculé. Il s'agit du la proportion de temps passé pour chaque activité au cours de la période d'observation, exprimée en pourcentage.

Un critère de « séquence d'activité » est calculé. A chaque changement de posture et/ou d'activité orale, le nombre de séquences d'activité est incrémenté d'une valeur. Ainsi une truie « couchée-immobile » qui passe à « couchée-activité orale non alimentaire » change de séquence. Ce critère mesure le nombre de fois où la truie change d'activité au cours de la période d'observation.

Une analyse de variance est réalisée avec le logiciel 'R'. Les élevages sont regroupés par grand système (BF ; RC ; DACsta ; DACdyn ; Tableau 1). Dans le cas du DAC, seules les truies qui ont consommé leur repas au cours des 2 heures qui suivent le début de la séquence alimentaire sont retenues soit 23 observations pour DACsta sur 60 observations potentielles (10 truies × 2 jours × 3 élevages) et 12 sur 40 pour DACdyn.

*Activité motrice des truies en groupes*

**Tableau 1 : Caractéristiques des élevages suivis**

|  | Bat-Flanc | | Réfectoire-courette | | | | DAC statique | | | DAC dynamique | |
|---|---|---|---|---|---|---|---|---|---|---|---|
| Elevages | 1 | 2 | 3 | 4 | 5 | 6 | 7 | 8 | 9 | 10 | 11 |
| Système alimentation | BF | BF | RC | RC | RC | RC | DAC | DAC | Selfifeeder | DAC | DAC |
| Sol | Cail. | Cail. | Lit. | Cail. | Lit. | Lit. | Cail. | Lit. | Cail. | Cail. | Lit. |
| Nb truies de l'élevage | 120 | 190 | 72 | 280 | 168 | 55 | 72 | 72 | 225 | 448 | 275 |
| Nb groupes/bande | 2 ou 3 | 4 | 4 | 2 | 3 | 1 | 1 | 1 | 1 | Groupe constitué de plusieurs bandes | |
| Nb truies/groupes | 6 à 8 | 6 à 7 | 5 à 6 | 12 | 8 à 10 | 18 | 24 | 24 | 45 | 280 | 220 |
| Surface de la case (m²) | 14,6 | 15,9 | 22,5 | 37,8 | 36,2 | 60 | 61,2 | 88,7 | 91 | 403 | 490 |
| Repas | 2 à 3 repas par jour à heure fixe | | | | | | Heure du repas de la truie selon ordre de passage au DAC- Séquence alimentaire | | | | |
| Observations comportement | 7h20-13h20 | 7h55-13h55 | 7h35-13h35 | 6h30-12h30 | 7h55-13h55 | 7h40-13h40 | 7h30-13h30 | 7h55-13h55 | 7h10-13h10 | 15h00-21h00 | 16h30-22h30 |

## 2. Résultats

### 2.1. Deux journées d'observation

Le comportement des truies est répétable au cours des deux journées successives d'observation. L'effet de la journée d'observation n'est pas significatif statistiquement sur les variables 'nombre de séquences d'activité' et 'activité orales non alimentaires' 'distance parcourue'. Une tendance (P<0,1) est obtenue pour les postures debout et couché.

### 2.2. Effet du type de sol

L'échantillon comporte 6 élevages où les truies sont logées sur caillebotis intégral et 5 où elles sont sur litière paillée. Le type de sol n'a pas d'effet sur les postures debout et couché, ni sur les distances parcourues ou l'inactivité.

En revanche, les activités orales des truies sont fortement liées à la présence ou non de paille. Sur caillebotis, 31% des observations sont consacrées à des activités orales non alimentaires. En présence de paille, la fréquence des activités orales non alimentaires est deux fois plus faible qu'en absence de litière (14% des observations), remplacée par une activité de manipulation de la paille (13,4% des observations).

### 2.3. Distance parcourue

La distance parcourue par la truie au cours de la période d'observation dépend avant tout de la surface de la case disponible. Logée avec un bat-flanc caillebotis, la surface de la case de 15 m² est réduite. Les truies parcourent en moyenne 50 m au cours des six heures d'observation. A l'opposé, les truies logées en DAC dynamique parcourent une distance 7 fois plus importante au cours de la période d'observation. La distance moyenne parcourue est de 362 m. Les truies logées aux réfectoire-courette et DAC statique occupent une position intermédiaire.

La figure 1 illustre la distance moyenne parcourue par les truies dans les 11 élevages observés. On observe que plus les surfaces offertes sont élevées, plus la distance moyenne parcourue est grande. Compte tenu de la configuration des salles, une truie logée au DAC dynamique doit parcourir une cinquantaine de mètres par jour ne serait-ce que pour aller boire et manger. Cette valeur est du même ordre de grandeur que la distance totale parcourue en moyenne par une truie au bat-flanc sur les 6 heures d'observation.

*Activité motrice des truies en groupes*

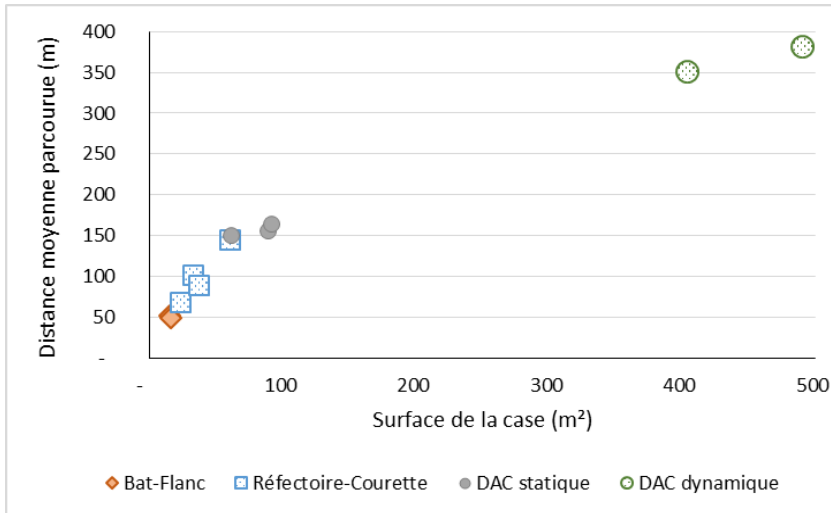

**Figure 1 :** distance moyenne parcourue dans les 11 élevages

**Tableau 2 : activité des truies en groupes selon les modes de logement**

| Groupe | Bat-flancs | Réfectoire-courette | DAC statique | DAC dynamique | ETR | Effet statistique |
|---|---|---|---|---|---|---|
| Nombre d'élevages | 2 | 4 | 3 | 2 | | |
| % Couché | 72,4 c | 56,8 b | 43,1 a | 41,9 a | 14,98 | *** |
| % Debout | 26,5 c | 39,9 b | 53,1 a | 55,7 a | 15,59 | *** |
| % Assis | 1,07 a | 3,2 a | 3,8 a | 2,3 a | 5,29 | NS |
| Distance (mètres) | 50 d | 101 c | 156 b | 362 a | 62,48 | *** |
| % Déplacement | 2,9 b | 3,9 b | 5,9 a | 8,4 a | 2,93 | *** |
| Nombre séquences d'activité | 12,6 c | 15,9 a | 27,6 b | 18,6 a | 5,04 | *** |
| % Activités orales non-alimentaires | 18,8 b | 25,6 ab | 29,8 ab | 36,7 a | 18,94 | * |
| % Inactivité | 75 b | 54,4 a | 51,5 a | 43,9 a | 15,88 | *** |
| % Alimentation | 2,2 a | 3,3 b | 6,2 c | 3,7 ab | 2,06 | *** |
| % Manipulation paille | | 11,6 a | 3,9 b | 6,7 ab | 9,92 | *** |
| % Contacts | 0,9 a | 0,97 a | 2,6 b | 0,37 a | 1,8 | *** |

### 2.1. Variabilité entre les truies

La Figure 2 illustre la dispersion observée sur les distances parcourues au cours de la période d'observation. Dans le système bat-flanc (BF), l'écart interquartile est de 20 m. En revanche, pour les systèmes réfectoire-courette (RC), DAC stable (DACsta) et DAC dynamique (DACdyn) il s'élève à 71 m, 47 m et 255 m, respectivement. Les distances les plus faibles s'observent en bat-flanc et réfectoire courette où certaines truies se contentent de se lever et se déplacer uniquement pour le repas (voir parcours P7). Au DAC dynamique, la distance minimale parcourue est de 173 m, une truie ayant parcouru jusqu'à 716 m au cours des 6 heures d'observation (parcours P3). L'ensemble des données liées à l'activité physique des truies (debout, déplacement) montrent le même type de variabilité entre modes de logement.

*Activité motrice des truies en groupes*

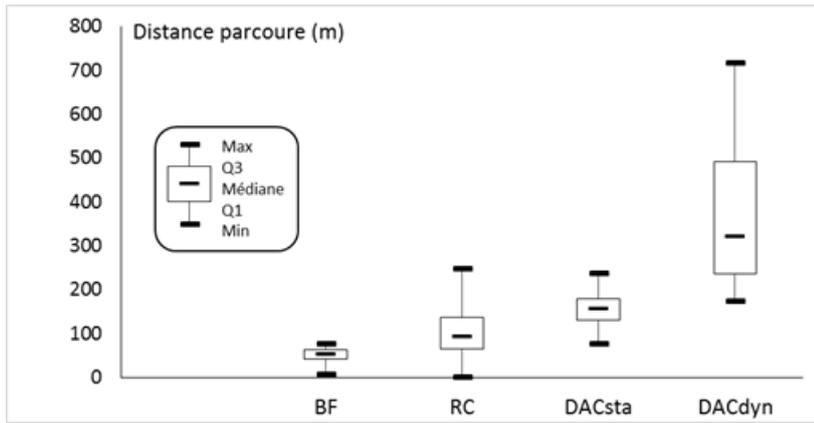

**Figure 2** : Dispersion de la distance parcourue (m) entre les truies et les systèmes de logement.

**Figure 3** : exemples de parcours des truies au cours de 6 heures d'observation

*(Ces exemples ne correspondent pas à un parcours type dans chacun des systèmes mais illustrent la diversité des situations)*

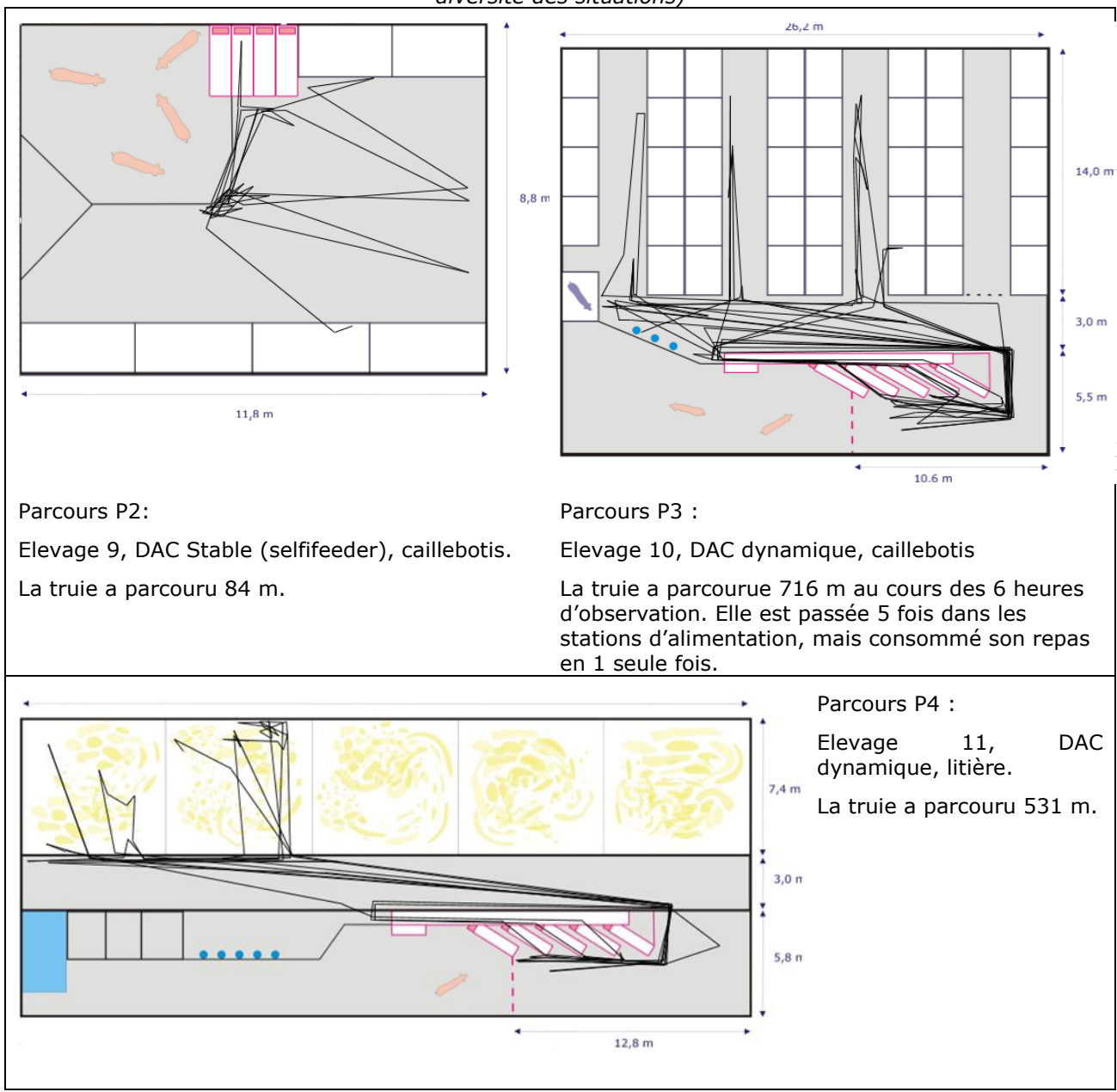

Parcours P2:

Elevage 9, DAC Stable (selfifeeder), caillebotis.

La truie a parcouru 84 m.

Parcours P3 :

Elevage 10, DAC dynamique, caillebotis

La truie a parcourue 716 m au cours des 6 heures d'observation. Elle est passée 5 fois dans les stations d'alimentation, mais consommé son repas en 1 seule fois.

Parcours P4 :

Elevage 11, DAC dynamique, litière.

La truie a parcouru 531 m.

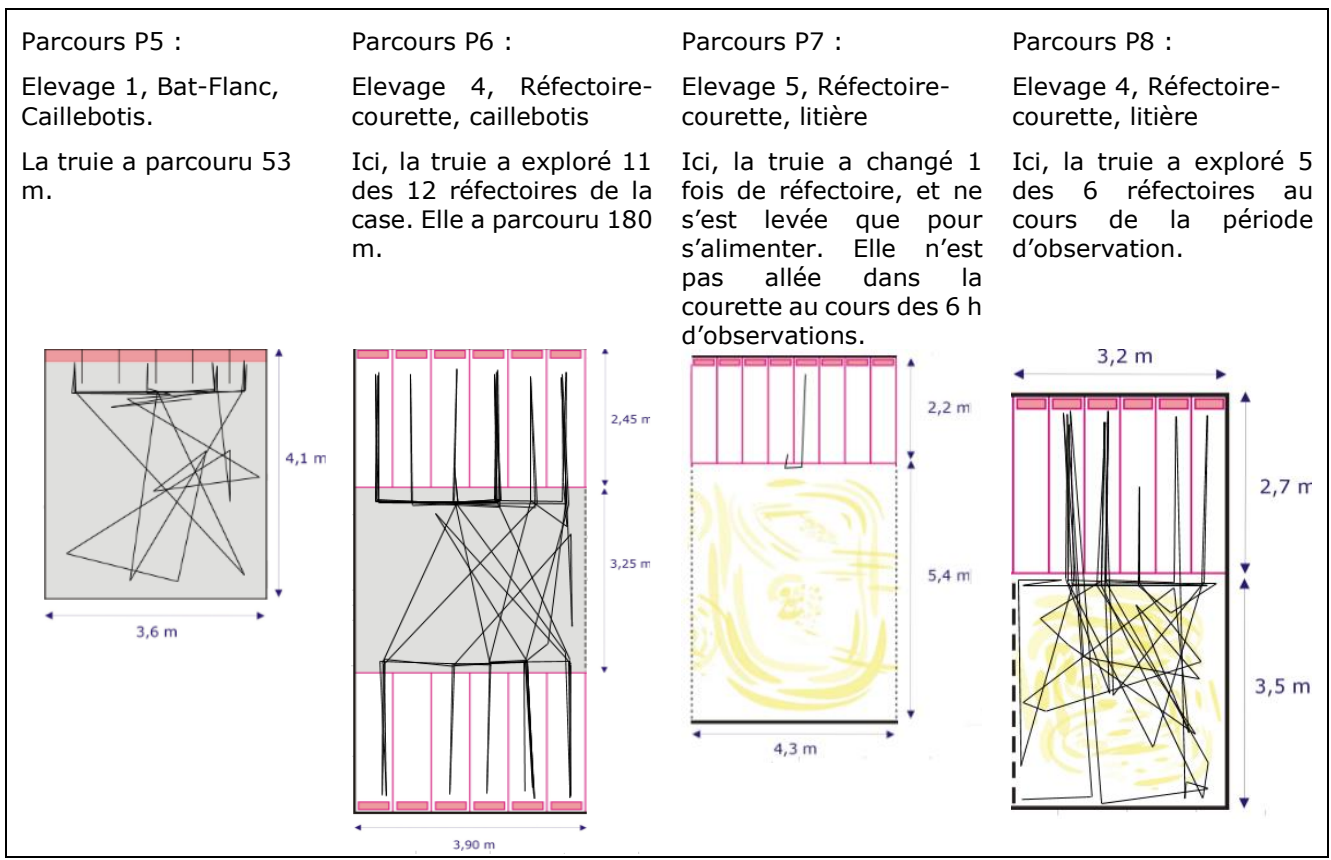

Parcours P5 :

Elevage 1, Bat-Flanc, Caillebotis.

La truie a parcouru 53 m.

Parcours P6 :

Elevage 4, Réfectoire-courette, caillebotis

Ici, la truie a exploré 11 des 12 réfectoires de la case. Elle a parcouru 180 m.

Parcours P7 :

Elevage 5, Réfectoire-courette, litière

Ici, la truie a changé 1 fois de réfectoire, et ne s'est levée que pour s'alimenter. Elle n'est pas allée dans la courette au cours des 6 h d'observations.

Parcours P8 :

Elevage 4, Réfectoire-courette, litière

Ici, la truie a exploré 5 des 6 réfectoires au cours de la période d'observation.

## 3. Discussion et Applications pratiques

Ce travail d'observations comportementales montre clairement les différences d'activité physique et motrice des truies selon le mode de logement en groupes.

Les truies alimentées en petites cases avec bat-flanc sont les moins actives. L'activité est concentrée autour des repas distribués en 2 ou 3 séquences. C'est également le cas avec les systèmes réfectoire-courette où les repas sont distribués à heure fixe. Certaines truies restent isolées dans leur réfectoire durant la période d'observation.

L'activité suit également les repas pour les truies alimentées au DAC, en groupe stable et dynamique. Des enregistrements video réalisées sur 24 heures dans 2 élevages de l'étude le montrent. En revanche, la période d'alimentation est beaucoup plus étalée dans le temps. Au DAC stable, les séquences alimentaires démarraient le matin et nous pouvions suivre aisément l'activité des truies retenues. En revanche, au DAC dynamique, les séquences alimentaires démarraient en fin d'après-midi pour se poursuivre pendant toute la nuit. Dans ces élevages, une séquence d'activité est généralement observée également le matin, liée au début de la période diurne, aux activités de l'éleveur (surveillance, tri des animaux, raclage des déjections..). Ce temps d'activité n'est pas pris en compte dans notre étude. La période en fin de matinée et début d'après-midi, située entre l'activité matinale et la séquence alimentaire du début de soirée, est la plus calme. Ceci a pu amener à minimiser les fréquences d'activités et les distances parcourues en DACdyn.

### 3.1. Type de sol

Les truies élevées sur litière orientent une part importante de leur activité orale vers la paille. Cette paille est fouillée, manipulée, ingérée parfois. En revanche, sur caillebotis les comportements sont principalement oraux non-alimentaires (mâchonnement à vide essentiellement) ou orientés vers des objets manipulables (des chaines étant systématiquement présentes dans les élevages).

La présence ou non de paille n'a pas d'effet statistique sur l'activité motrice des truies, en termes de temps passé debout et de distance parcourue. Sur ces critères, il semble que la dimension de la case joue davantage.

### 3.2. Besoins alimentaires liés à l'activité physique

Les besoins d'entretien de la truie sont doublés lorsqu'elle est debout comparativement à la position couchée (Noblet et al., 1994). L'animal est dit « à l'entretien » quand il ne perd ni ne gagne de poids, que son niveau d'activité est faible et qu'il est placé à la thermoneutralité.

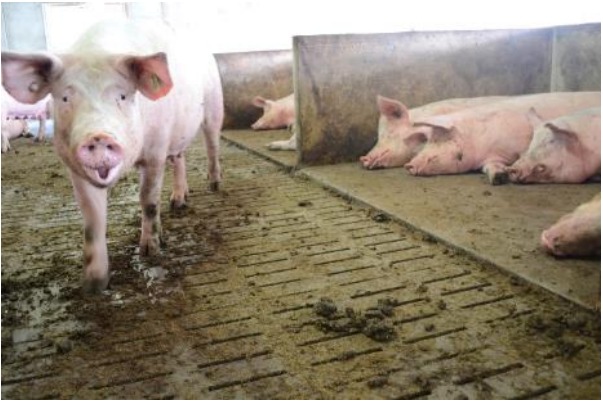

*Photo 2: Par rapport à la position couchée, le besoin d'entretien de la truie est doublé lorsqu'elle est debout*

Les données liées à l'activité physique mesurées dans la présente étude soulignent la nécessité de différencier les besoins liés à l'activité selon le mode de logement en apportant davantage d'aliment aux truies logées dans de grandes cases.

Le second paramètre est la variabilité de l'activité entre les truies, plus importante au DAC qu'au bat-flanc. Dans les systèmes de type bat-flanc et réfectoire-courette, l'aliment est généralement apporté sous forme de repas à l'ensemble du groupe. Compte tenu de la faible variabilité d'activité entre les individus du groupe, l'alimentation non individualisée est moins gênante.

En revanche, au DAC, notamment dans les groupes dynamiques, les différences entre les truies sont importantes. Le mode de distribution permet cependant d'ajuster la ration au besoin individuel de chaque truie. La difficulté majeure reste d'identifier au sein du groupe le niveau d'activité de chaque individu.

### 3.3. Déplacement et problèmes d'aplombs

Peu de problèmes locomoteurs ont été observés dans les élevages retenus pour cette étude et aucune truie suivie ne présentait de boiterie. Les problèmes locomoteurs sont plus importants depuis que les truies sont en groupes, notamment sur caillebotis. Les causes sont multiples (Anil et al., 2007), liées au logement, à la ventilation, mais également aux apports alimentaires (van Riet et al., 2013).

Des observations récentes en élevages montrent que le système DAC est favorable aux boiteries (Cador et al., 2013). Ceci pourrait s'expliquer en partie par l'activité motrice des truies plus élevée dans ce système.

On peut émettre l'hypothèse que dans les systèmes DAC, les truies sont forcées de se déplacer ne serait-ce que pour s'alimenter et s'abreuver. Avec une cinquantaine de mètre à parcourir, la distance minimale que doivent parcourir les truies au DAC dynamique est du même ordre de grandeur à la distance moyenne parcourue par les truies logées au bat-flanc. Au DAC, la truie n'a pas moyen de s'isoler même si elle présente un problème locomoteur qui pourrait se résorber avec un repos. Une période de repos est cependant favorable à l'amélioration des problèmes locomoteurs. Ainsi dans 5 élevages, la période de maternité permet d'améliorer l'état des aplombs des truies (Caille, 2013). Il convient donc, en situation au DAC de repérer précocement les truies qui présentent des problèmes locomoteurs (boiteries) et de les isoler dans des infirmeries.

## 4. Conclusion

Le comportement des truies est différent selon le mode de logement. Plus la surface de la case est importante, plus les truies se déplacent et parcourent une distance élevée. La distance parcourue au DAC dynamique au cours des 6 heures d'observation est ainsi 7,2 fois supérieure à celle observée dans un logement de type bat-flanc. Le type de sol (caillebotis/litière) impacte le comportement oral des truies.

Ce travail suggère de prendre en compte l'activité physique des truies différemment selon les modes de logement pour l'évaluation du besoin nutritionnel. Une évaluation de l'activité motrice sur une durée de 24h permettrait de préciser davantage ce paramètre.

## 5. Pour plus d'informations

Contact : Yannick Ramonet

Pôle porc-aviculture des Chambres d'agriculture de Bretagne, Plérin

Tél.: 02 96 79 21 90

Mail : yannick.ramonet@bretagne.chambagri.fr

## 6. Références bibliographiques

**Assessment of the motor activity of group-housed sows in commercial farms**

The objective of this study was to specify the level of motor activity of pregnant sows housed in groups in different housing systems. Eleven commercial farms were selected for this study. Four housing systems were represented: small groups of five to seven sows (SG), free access stalls (FS) with exercise area, electronic sow feeder with a stable group (ESFsta) or a dynamic group (ESFdyn). Ten sows in mid-gestation were observed in each farm. The observations of motor activity were made for 6 hours at the first meal or at the start of the feeding sequence, two consecutive days and at regular intervals of 4 minutes. The results show that the motor activity of group-housed sows depends on the housing system. The activity is higher with the ESFdyn system (standing: 55.7%), sows are less active in the SG system (standing: 26.5%), and FS system is intermediate. The distance traveled by sows in ESF system is linked to a larger area available. Thus, sows travel an average of 362 m ± 167 m in the ESFdyn system with an average available surface of 446 m² whereas sows in small groups travel 50 m ± 15 m for 15 m² available.

**Comment citer ce document ?**

Yannick Ramonet, Anaïs Tertre, 2013 Activité motrice des truies en groupes dans les différents systèmes de logement. Rapport d'étude. Chambres d'agriculture de Bretagne, 8 pages.

**Ce travail a donné lieu à 2 publications**

- Ramonet Y., Tertre A., 2014. De 50 à 500 mètres par jour. Chaque truie à son rythme. Tech-Porc, n°15 : 20-22.
- Tertre A., Ramonet Y., 2014. Evaluation de l'activité motrice des truies en groupes en élevages de production. Journée de la Recherche Porcine, 46 : 267-268.

**Mots-clés : truie, comportement, activité motrice**

*Activité motrice des truies en groupes*

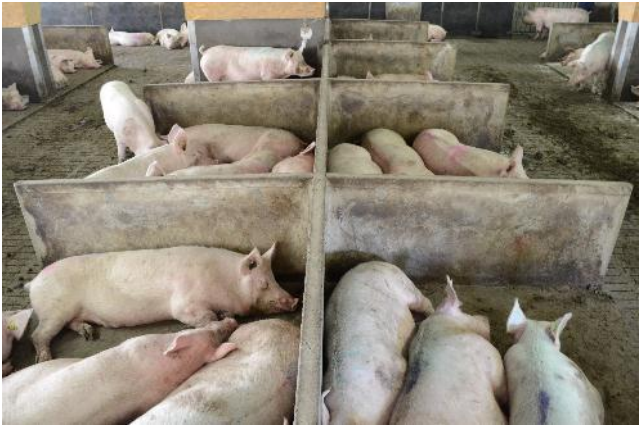 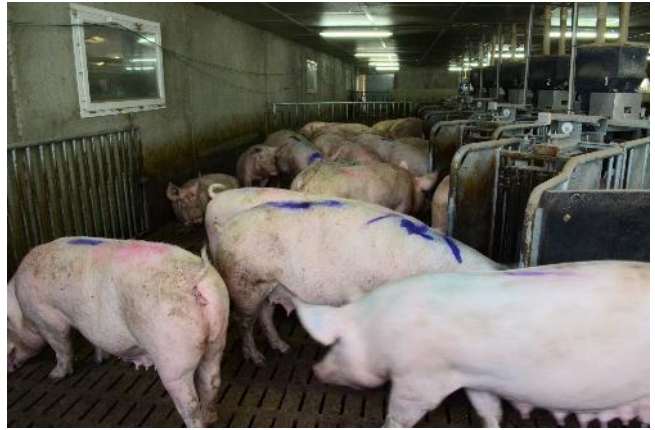

*L'activité des truies logées avec un DAC est également dépendante du début du cycle alimentaire. L'impression que l'observateur peut avoir en milieu d'après-midi lorsque les animaux sont au repos ne reflète pas l'activité nocturne des truies lorsque la séquence alimentaire débute vers 17h00-18h00.*